\documentclass{article}

\usepackage{amsmath}
\usepackage[hidelinks]{hyperref}
\usepackage[super]{nth}
\usepackage[nottoc,numbib]{tocbibind}
\usepackage{cite}
\usepackage{siunitx}
\usepackage[labelfont=bf]{caption}
\usepackage{float}
\usepackage[margin=2in]{geometry}
\usepackage{multirow}
\usepackage{rotating}
\usepackage{graphicx}
\usepackage{authblk}
\usepackage{parskip}

\newcommand{\finf}{\mathcal{I}}
\newcommand{\bs}[1]{\boldsymbol{#1}}
\newcommand{\email}[1]{\href{mailto:#1}{#1}}

\title{Spatial Information in Large-Scale Neural Recordings}
\author[1,*]{Thaddeus~ R.~Cybulski}
\author[1,*]{Joshua~I.~Glaser}
\author[2,3]{Adam~H.~Marblestone}
\author[4]{Bradley~M.~Zamft}
\author[5,6,7]{Edward~S.~Boyden}
\author[2,3,4]{George~M.~Church}
\author[1,8,9]{Konrad~P.~Kording}
\affil[1]{Department of Physical Medicine and Rehabilitation, Northwestern University and Rehabilitation Institute of Chicago, Chicago, Illinois, USA}
\affil[2]{Biophysics Program, Harvard University, Boston, Massachusetts, USA}
\affil[3]{Wyss Institute, Harvard University, Boston, Massachusetts, USA}
\affil[4]{Department of Genetics, Harvard Medical School, Boston, Massachusetts, USA}
\affil[5]{Media Lab, Massachusetts Institute of Technology, Cambridge, Massachusetts, USA}
\affil[6]{Department of Biological Engineering, Massachusetts Institute of Technology, Cambridge, Massachusetts, USA}
\affil[7]{McGovern Institute, Massachusetts Institute of Technology, Cambridge, Massachusetts, USA}
\affil[8]{Department of Physiology, Northwestern University, Chicago, Illinois, USA}
\affil[9]{Department of Applied Mathematics, Northwestern University, Chicago, Illinois, USA}
\affil[*]{\textnormal{Authors contributed equally to this work. Direct correspondence to \email{cyb@northwestern.edu} and \email{j-glaser@u.northwestern.edu}}}

\date{}

\begin{document}

\maketitle

\begin{abstract}
To record from a given neuron, a recording technology must be able to separate the activity of that neuron from the activity of its neighbors. Here, we develop a Fisher information based framework to determine the conditions under which this is feasible for a given technology. This framework combines measurable point spread functions with measurable noise distributions to produce theoretical bounds on the precision with which a recording technology can localize neural activities. If there is sufficient information to uniquely localize neural activities, then a technology will, from an information theoretic perspective, be able to record from these neurons. We (1) describe this framework, and (2) demonstrate its application in model experiments. This method generalizes to many recording devices that resolve objects in space and should be useful in the design of next-generation scalable neural recording systems.
\end{abstract}

\tableofcontents

\section{Introduction}
A concerted effort is underway to develop technologies for recording simultaneously from a large fraction of neurons in a brain~ \cite{alivisatos2013,marblestone2013}. For a technology to reach the goal of large-scale recording, it must gather sufficient information from each neuron to determine its activity. This suggests that neural recording methodologies should be evaluated and compared on information theoretic grounds. Still, no widely applicable framework has been presented that would quantify the amount of information large-scale neural recording architectures are able to capture. Such a framework promises to be useful when we want to compare the prospects of new recording technologies.

A neural recording technology can be judged by its ability to isolate signals from individual neurons. One common method of differentiating between signals from different neurons is through the neurons' locations: if the recording technique can determine the signal sources are sufficiently far apart (by signal amplitude or other methods), then the signals likely came from different neurons. One can quantify this ability to spatially differentiate neurons using Fisher information, which measures how much information a random variable (e.g. a signal on a detector) contains about a parameter of interest (e.g. where the signal originated). Fisher information can be used to determine the optimal precision with which the parameter of interest (the neural location) can be estimated.\footnote{Fisher information is a theoretical calculation that determines the best a technology can do -- signal separation techniques (e.g. \cite{mukamel2009}) are generally required to approach this optimum.} By calculating the Fisher information a technology carries about sources it records, one can determine how precisely neural locations can be estimated using this technology, and thus whether the neural activities can be distinguished in space.

Determining the Fisher information content of a sensing system allows determining the informatic limits of a technology in a given situation. These informatic limits, in turn, can guide technology design. For example, by quantifying the information content of an electrode array as a function of the spacing between electrodes, one could determine the spacing necessary to distinguish neural activities. Similarly, one can compare the information content of several optical recording approaches to determine the optimal technology for a given experiment.

Here we develop a Fisher information-based framework that characterizes neural recording technologies based on their abilities to distinguish activities from multiple neurons. We apply this framework to models of neural recording techniques, describe how the Fisher information scales with respect to recording geometries and other parameters, and demonstrate how this framework could be utilized to optimize experimental design.We demonstrate the utility of a Fisher information-based evaluation of neural recording technologies, which may inform the design and development of next-generation recording techniques.

\section{Framework}
\subsection{Localization and Resolution}
A fundamental concern in neural recording is “localization,” the ability to accurately estimate the location of origin of neural activity. Localization is a primary method of determining the identity of an active neuron.

\begin{figure}[H]
\centering
\includegraphics[scale=0.45]{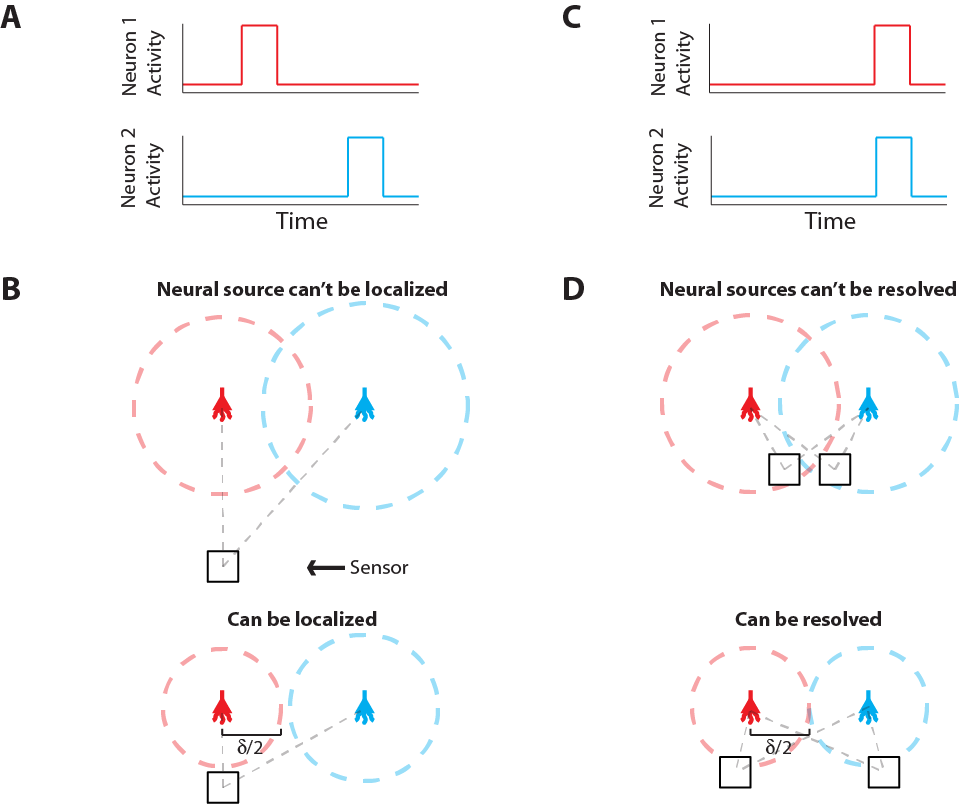}
\caption[Localization and Resolution]{\textbf{Localization and Resolution.} \textbf{(A)} In many behavioral states, neural systems have sparse activity, in which neighboring neurons (red and blue) are not active at the same time. In this scenario of single-source resolution, one neuron must be localized at a given time. Panel \textbf{B} looks at this scenario. \textbf{(B)} Two neighboring neurons are shown a distance $\delta$ away from each other. Dotted lines indicate regions where we are confident about the source of a signal, i.e. we have a sufficient amount of information regarding that signal's location. The signals from the two neurons are recorded by the sensor at different times and do not interfere with each other. When a neuron cannot be localized effectively, i.e. there is not sufficient Fisher information, it is because the signal from that neuron was not strong enough to overcome noise. \textbf{(C)} Sometimes, neighboring neurons are simultaneously active. In this scenario of differential resolution, both neurons must be localized at a given time. Panel \textbf{D} looks at this scenario. \textbf{(D)} Same as \textbf{B}, except two sensors are necessary for differential resolution. When both sensors record similar signals, i.e. when there is large mutual information regarding the two neurons' activities, it is difficult to resolve the neurons.} \label{fig:1}
\end{figure}

The problem of establishing neural locations can be split into two separate regimes. One regime is when an active neuron has no active neighbors (\autoref{fig:1}A). In this state, we are chiefly concerned with the ability to attribute the signal to the correct neuron (single-source resolution~\cite{dekker1997}). This can be done by accurately localizing one activity at a given time on a background of noise (\autoref{fig:1}B). The other regime is when two neighboring neurons are simultaneously active (\autoref{fig:1}C). In this state, we are chiefly concerned with the ability to differentiate the two neurons, i.e. are there two “clearly distinguished” or one “blurred” neuron (differential resolution~\cite{dekker1997}). This can be done by simultaneously localizing the activities of both neurons accurately (\autoref{fig:1}D).\footnote{While we have been discussing differentiating neurons, the framework itself differentiates between point sources. In this paper, we make the assumption that separate point sources belong to separate neurons. In reality, it is possible that there could be separate signals from the cell body and dendrites that are perceived as different sources. These can be united using additional information (e.g. anatomical imaging or simultaneous activity).}

Fisher information can be used to determine whether both scenarios are theoretically possible for a given technology. Here we treat both of these scenarios: first by calculating the Fisher information a sensing aparatus has about the location of a single neuron, then expanding this framework to treat location parameters of multiple neurons. We address localization and resolution in the theoretical limit where the point spread function (PSF) is known, in order to study the limiting effects of neuronal and sensor noise on localization precision.\footnote{There exists a family of deconvolution techniques that estimate the PSF and use it to obtain a more accurate representation of the original signal (e.g.~\cite{onodera1998,colak1997,yan2008,broxton2013}). In theory, with sufficient samples and knowledge of the PSF, one could obtain a perfect representation of a sparse signal in the absence of noise. This is not the case in practice, as signals are not only modified reversibly by PSFs, but are modified irreversibly by noise on neurons and detectors (e.g.~\cite{shahram2004,shahram2005}). In the presence of noise and other aberrations, it thus becomes difficult to isolate individual sources using deconvolution techniques, even when the PSF is known. Thus, it is interesting to determine the isolated effects of noise on recording methods. Moreover, as this Fisher information framework gives optimal bounds on precision with a known PSF, it can be used to determine how close to optimal a deconvolution algorithm performs.}

Regardless of the number of neurons and sensors we are treating, Fisher information gives us a metric with which to evaluate a recording technology. “Spatial information,” the amount of information regarding the location of a source (i.e., a quantitative measure of localization ability), can be used to determine whether it is possible to correctly attribute an activity to its source (or multiple activities to multiple sources). In order to know the identity of a source, we must be confident about the location of origin of the activity with a positional error less than $\delta$, where $\delta$ is the distance from one neuron to another (\autoref{fig:1}B\&D). In terms of Fisher information, if we have sufficient information to locate the source of activity with a precision $\delta$, we can assign that activity to a single neuron that occupies that location.

\subsection{Fisher Information: General Principles}
Fisher information is a metric that measures the information a random variable has about a parameter, and can be used to determine how well that parameter can be estimated. More precisely, Fisher information, $\finf(\theta )$ is a measure of the information a random variable $X$, with distribution $f(X;\theta)$ parameterized by $\theta$, contains about the parameter $\theta$~\cite{kullback1997}:

\begin{equation}
\finf(\theta ) = \operatorname{E} \left[ {\left. {{{\left( {\frac{\partial }{{\partial \theta }}\log f(X;\theta )} \right)}^2}} \right|\theta } \right] = \int {{{\left( {\frac{\partial }{{\partial \theta }}\log f(x;\theta )} \right)}^2}} f(x;\theta )\;dx
\end{equation}
Intuitively, the more $X$ changes for a given change in $\theta$ (the greater the magnitude of $\frac{\partial }{{\partial \theta}} f(X;\theta )$ ), the more information you will know about $\theta$ by observing $X$.

More generally, the Fisher Information a random variable $X$ has about a parameter vector $\bs{\theta}$ with $k$ elements $[\theta_1 \cdots \theta_k]$ can be represented by a $k$ x $k$ matrix with elements:

\begin{equation}
{\left( {\finf(\bs{\theta})} \right)_{ij}} = \operatorname{E} \left[ {\left. {\left( {\frac{\partial }{{\partial {{\theta}_i}}}\log f(X;\bs{\theta})} \right)\left( {\frac{\partial }{{\partial {{\theta}_j}}}\log f(X;\bs{\theta})} \right)} \right|\bs{\theta} } \right]
\end{equation}
The elements of this matrix represent the information contained in a sample about a pair of parameters.

\subsection{Cramer-Rao Bounds}
The optimal precision with which the parameter, $\theta$, can be estimated is inversely related to the Fisher information contained about that parameter. More precisely, the variance of an unbiased estimator of a parameter is lower bounded by the Cramer-Rao bound (CRB)~\cite{cramer1946}:

\begin{equation}
\operatorname{Var} \left[ \bs{{\hat \theta}}_{i} \right] \geq \left[\finf\left(\bs{\theta}\right)^{-1}\right]_{ii}
\end{equation}

An important implication of this is that the CRB on $\theta_i$ not only depends on the information $X$ contains about $\theta_i$, but how similar $\theta_i$'s effect on $X$ is to the rest of the elements of $\bs{\theta}$. An off-diagonal term ${\left( {\finf(\bs{\theta})} \right)_{ij}}$ with large magnitude means that the parameters $i$ and $j$ are strongly correlated (or anti-correlated) in terms of their input on $X$. This will increase the CRB on estimating parameters $i$ and $j$.

\subsection{Independence and Summation}
If two observations $X_1$ and $X_2$ are independently affected by $\bs{\theta}$, then the two Fisher information matrices about $\bs{\theta}$ can be summed, as could be expected by the implications of independence on sample variance. This property allows us to easily apply our framework to situations with multiple samples, either by multiple sensors or multiple time points.

In the following sections, we will apply the above properties of Fisher information and CRBs to develop a framework for determining how precisely the location of neural activities can be estimated, and thus whether they can be distinguished. Note that, while we will describe the ability to distinguish neurons solely using spatial information, additional sources of information can be used, e.g., temporal information in optical~\cite{pnevmatikakis} and electrical recordings~\cite{lewicki1998} (see \textit{\nameref{sec:discussion}}).

\subsection{Fisher Information: Single-Source Resolution}

We first examine the situation where a single active source of some known intensity must be localized using an ensemble of sensors.\footnote{Activity in neural systems is often sparse~\cite{barth2012,denman2013,cohen2010,cohen2011,bair2001}; this simplified scenario may be a useful model of neural systems.} Here we observe a random variable, $X$, the value recorded at some sensor (e.g. in Volts). $f(X;\bs{\theta})$ then is the distribution of sensor values from repeated recordings of a neuron parametrized by $\bs{\theta}$. $\bs{\theta}$ is a vector representing spatial (and other) parameters that  characterize the neural signal. This resulting distribution $f(X;\bs{\theta})$ reflects both intrinsic variance of a neural signal as well as extrinsic factors such as other neurons and noise.

Here, Fisher information, $\finf{\left(\bs{\theta} \right)}$, measures how much the distribution of recorded sensor values $f(X;\bs{\theta})$ tells us about the location of a signal's origin (\autoref{fig:2}D). Intuitively, if a change in the signal origin's location would cause a large change in the recorded signal, then there will be a large amount of information about the location. However, if a change in the origin of the neural signal does not affect the recorded signal, there will be little information about the location of the neuron.

The CRB for a given parameter ${\theta}_i$ will tell us how precisely that location parameter can be estimated from the signal intensity. Assuming an unbiased estimator (the average estimate will be the true location), the best possible variance of the estimate is $[\finf\left(\bs{\theta}\right)^{-1}]_{ii}$. If we want to be confident that the estimated location of a given neuron's activity is within $\delta/2$ of its true location, as in \autoref{fig:1}, the CRB on the estimate of distance must be less than $\left(\delta/4 \right)^2$).\footnote{95\% confidence under Gaussian assumptions}

Without assuming any prior knowledge, at least $k$ variables are required to estimate $k$ parameters, as the system is underconstrained with smaller numbers of samples. In our case, we need multiple sensors in order to estimate a neuron's location. If the sensors have independent noise -- an assumption we use in our demonstrations -- the information matrices can be summed (See \textit{Independence and Summation}).

\subsection{Fisher Information: Differential Resolution}

In the scenario of multiple neurons acting simultaneously, we are interested in using signals recorded from an ensemble of sensors to estimate the location parameters of each neuron. That is, $\bs{\theta}$ now represents the location parameters of all neurons in the system, and $f(X;\bs{\theta})$ represents the distribution of signal intensities on a sensor given all of the neurons in the system. We can then construct a Fisher information matrix to determine the precision with which each parameter can be estimated. If each sensor recording is affected by $n$ neurons, each with $k$ parameters, the Fisher information matrix will be $nk \times nk$. The CRB calculated in this scenario will be most applicable to determining whether technologies are able to effectively record from a population of neurons.

\subsection{Point Spread Functions and Signal Intensity Distributions}
\begin{figure}[H]
\includegraphics[width=\linewidth]{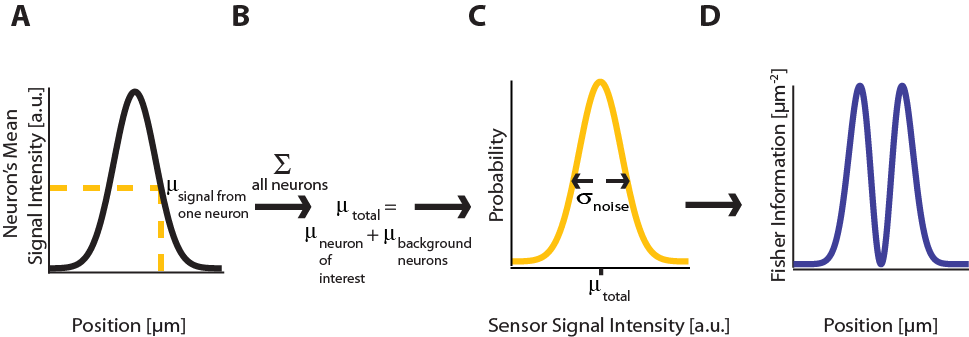}
\caption[Fisher Information]{\textbf{Fisher Information.} \textbf{(A)} A signal on sensor $i$ from a neuron $j$ at a particular location has a mean intensity, defined by a recording method's point spread function and the intensity of the signal from the active neuron. We here plot this mean signal intensity as a function of one position parameter. \textbf{(B)} The mean total signal on a sensor, $\mu_{total}$, is the sum of the signals from every neuron. \textbf{(C)} The distribution of intensities recorded on a sensor is a function of the total mean signal, $\mu_{total}$, and the variance of that signal, $\sigma^2_{noise}$, which can result from many different noise sources. \textbf{(D)} Fisher information can be derived from the distribution of signal intensity values on a sensor.}\label{fig:2}
\end{figure}

\begin{subequations}
To determine the spatial Fisher information, we must know the distribution of signals on a sensor given the location of the activity, $f(X;\bs{\theta})$. In this section, we derive the general form of $f(X;\bs{\theta})$ based on the PSF of a technology.

The signal measured by many recording systems is well approximated as a linear function of the signals from each neuron in a population~\cite{johnston1995,cremer2013}, i.e. the total sensor signal is the sum of the individual neural signals weighted by the magnitude of their individual effects on the sensor (\autoref{fig:2}A\&B). We thus only consider linear interactions; it should be noted that the Fisher information framework is also compatible with nonlinear interactions (e.g. sensor saturation). For $N$ neurons and $M$ sensors in a system, in the absence of noise, the signal on any particular sensor can therefore be described as:

\begin{equation}
{\mathbf{x = Wa + \epsilon}}
\end{equation}
where $\mathbf{x}$ is the vector of signals on sensors $[X_1,\cdots,X_M]$, $\epsilon$ is noise on the signal from neurons and sensors, and $\mathbf{a}$ is the vector of signals from neural activities, $[I_1,\cdots,I_N]^T$, e.g. the fluorescent signal produced due to neural activity in optical techniques or the voltage signal in electrical techniques. $\mathbf{W}$ is the matrix of PSFs:

\begin{equation}
{\mathbf{W}} = \left[ {\begin{array}{*{20}{c}}
{w({\bs{d}^{1,1}})}& \cdots &{w({\bs{d}^{1,N}})} \\ 
\vdots & \ddots & \vdots \\ 
{w({\bs{d}^{M,1}})}& \cdots &{w({\bs{d}^{M,N}})} 
\end{array}} \right]
\end{equation}
where $w$ is the PSF, which depends on the location of the neuron relative to the sensor and other parameters of a recording modality (e.g. light scattering). $\bs{d}^{i,j}$ is a vector that gives the location of neuron $j$ relative to sensor $i$. It has elements $[d^{i,j}_1 \cdots]$ that describe single location parameters of $\bs{d}^{i,j}$. 
\end{subequations}

\begin{subequations}
Combing Eqs. 4a and 4b, we can write the total signal on a sensor $i$ as

\begin{equation}
X_i={\sum\limits_j {{I_j}w({\bs{d}^{i,j}})}}+\epsilon
\end{equation}
We can write a function $f(X_i)$ that characterizes the distribution of signal intensities on a sensor. Here, we assume that the noise, $\epsilon$, can be approximated by a zero-mean Gaussian with variance $\sigma^2_{noise}$, so that:

\begin{equation}
\label{eq:normtog}
{f}(X_i;\bs{\theta}) = {\mathcal N}\left( {\sum\limits_j {{I_j}w({\bs{d}^{i,j}})} ,\sigma _{noise}^2} \right)
\end{equation}
where $\mathcal{N}(\mu,\sigma^2)$ signifies a normal distribution (\autoref{fig:2}C). $\bs{\theta}$ is the vector of parameters that we are estimating. It can include any $I_j$ and any elements of any $\bs{d}^{i,j}$.
This allows us to calculate the Fisher information in signal $X_i$ about location parameters of neurons using Eqs. 1 or 2 (Figure 2D).
\end{subequations}

It is important to note that, as long as they can be analytically described, all types of noise (of which there are many; see \textit{\nameref{sec:supinfo}} for further discussion) can be incorporated into this framework. This flexibility in noise sources makes this framework especially relevant for neural recording. 

\section{Framework Discussion}

Here we have described a framework to quantitatively approach the challenges of large-scale neural recording and determine the necessary experimental parameters for potential recording modalities. This framework extends previous work applying Fisher information to individual imaging techniques (e.g.~\cite{shahram2005,quirin2012,ober2004,aguet2005,winick1986,shahram2006,sanches2010,marengo2009,helstrom1969,mukamel2012,shechtman2014}) by considering a PSF and noise model  based on recording in neural tissue, and then using the CRB to establish signal separability. It is able to describe the information content of neural recording technologies that separate sources based on their locations. This information content can then be used to evaluate a technology's ability to separate sources. Such a framework promises to be useful in evaluating and comparing novel and established recording technologies.

Given this framework's reliance on signal modulation by PSFs, it neglects other ways that sources can be separated, such as color~\cite{hampel2011} or spike waveform. Some of this information could be made compatible with our framework via virtual recording channels, e.g. in time. While these types of non-spatial information are not considered here, they may be necessary to separate sources under certain recording situations, e.g. where the dendrites of one neuron produce a signal within the CRB of the cell body of another neuron. In an extreme case, proposed intracellular molecular recording devices have no spatial information, but could still effectively separate signals.~\cite{kording2011,zador2012} While spatial Fisher information is an attractive method of evaluating neural recording techniques, it is important to remember these limitations when considering “non-spatial” techniques.

In addition, the CRBs described here only consider unbiased estimators. That is, they only provide a lower bound on localization ability when there are no prior assumptions about neurons’ locations. It is possible to be more precise than the CRB if the estimator is biased (i.e. if assumptions are made about neurons’ locations, or neurons’ locations are constrained). There is work on Bayesian Cramer Rao Bounds~\cite{vantrees2004,dauwels2005} and bounds on parameter estimation with constraints~\cite{gorman1990,matson2006} that could be applied to better understand the capabilities of recording technologies.

This framework is particularly suited to the evaluation of novel techniques due to its general nature; it is applicable to any technique where a spatial PSF can be measured and the system's noise distribution can be either modeled or explicitly described. For instance, advanced optical techniques,~\cite{ahrens2013,prevedel2014} ultrasound, and MRI have all been proposed as potential large-scale neural recording techniques~\cite{marblestone2013,seo2013}. With a PSF describing how signals from different positions in the brain reach a sensor (some discussion in~\cite{prevedel2014,shin2009,jensen1991,qin2012,smith1998,engelbrecht2006}) and further quantification of recording noise, this framework could easily be applied to determine bounds on signal separability for those techniques.

Ultimately, the utility of this approach is dependent on the quality of PSFs and noise models we have. For some techniques, these are well-described (especially PSFs); for others, these are poorly understood. As models of neural recording techniques advance, the predictions of this technique will become more accurate.

\section{Demonstrations}
\label{sec:apps}
Here, we demonstrate the utility of the Fisher information framework for analysis of neural recording technologies. We provide demonstrations of the use of Fisher information in the cases of single-source and differential resolution. We first calculate the spatial Fisher information of a single source in simple recording setups for several model recording methods. We next demonstrate more realistic uses of the Fisher information framework using multiple neurons: optimal technology design and technology comparison.

\subsection{Assumptions}
For our demonstrations, we make several assumptions. First, we assume that all activity from the neuron of interest, including the noise, is part of the signal of interest. Thus, the total noise is a function of the sensor noise plus the noise of all neurons except for the neuron of interest. In order to create an accurate model of a neural recording technology, we must know how all sources of noise affect the recorded signal, and also the relation between the noise and the intensity of the neural activity. Because these are in general not known, we make further assumptions in our simulations.

In regards to neural activities, we assume that every active neuron has the same activity $I_0$, while non-active neurons have no activity, that the neuron of interest, $k$, is active at the moment we sample, and that other neurons are active at a uniform rate. We assume noise sources from neurons are independent, so that:

\begin{equation}
\sigma _{noise}^2 = \sum\limits_{j \ne k} {\sigma _j^2}
\end{equation}

There are many sources of noise, both on neurons and sensors, that could be included; these are discussed in the \textit{\nameref{sec:supinfo}}. For our demonstrations, we consider signal dependent noise that can arise from neurons and/or sensors. Specifically, for analytic simplicity, we only consider noise that has a standard deviation proportional to the mean signal: $\sigma _j^2 \propto I_0^2 \left(w({\bs{d}^{i,j}}) \right)^2 $. We use these simplifying assumptions so that the magnitudes of the fluorescence (optical) and waveform voltage (electrical) have no influence on the final information theory calculations (and the relationship between these magnitudes and the noise is not in general well understood). We emphasize that these simulation assumptions are implemented to simply demonstrate the use of this framework; more realistic outputs could be found using more complex, realistic noise models.

\subsection{Single Neuron Localization}
Here we calculate Fisher information of recording technologies using a single neuron and simple sensor arrangements as an illustration of our framework. We look at three technologies: (1) electrical recording, a traditional neural recording modality, (2) wide-field fluorescence microscopy, a traditional optical approach, and (3) two-photon microscopy, a modern optical approach. These examples are chosen for their relative simplicity and ability to illustrate the flexibility of a Fisher information approach to modeling neural recording.

For any technology, the aim is for there to be, across all sensors, sufficient information about every location in the brain in order to identify a neuron firing in that location. Thus for an individual sensor, it can be better to have sufficient (enough to identify a neuron, as in \autoref{fig:1}) information spread over a large area than excessive information about a small area. This suggests that experimental designs could be modified to get sufficient information for the required task. For example, an optical technology may have extra information at low depths, but insufficient information at large depths. In this case, the PSF could be modulated (e.g.~\cite{quirin2012}) to decrease low-depth information (making those images blurrier), while increasing high-depth information.

\subsection{Electrical Sensing}
\begin{figure}[H]
\centering
\includegraphics[scale=0.5]{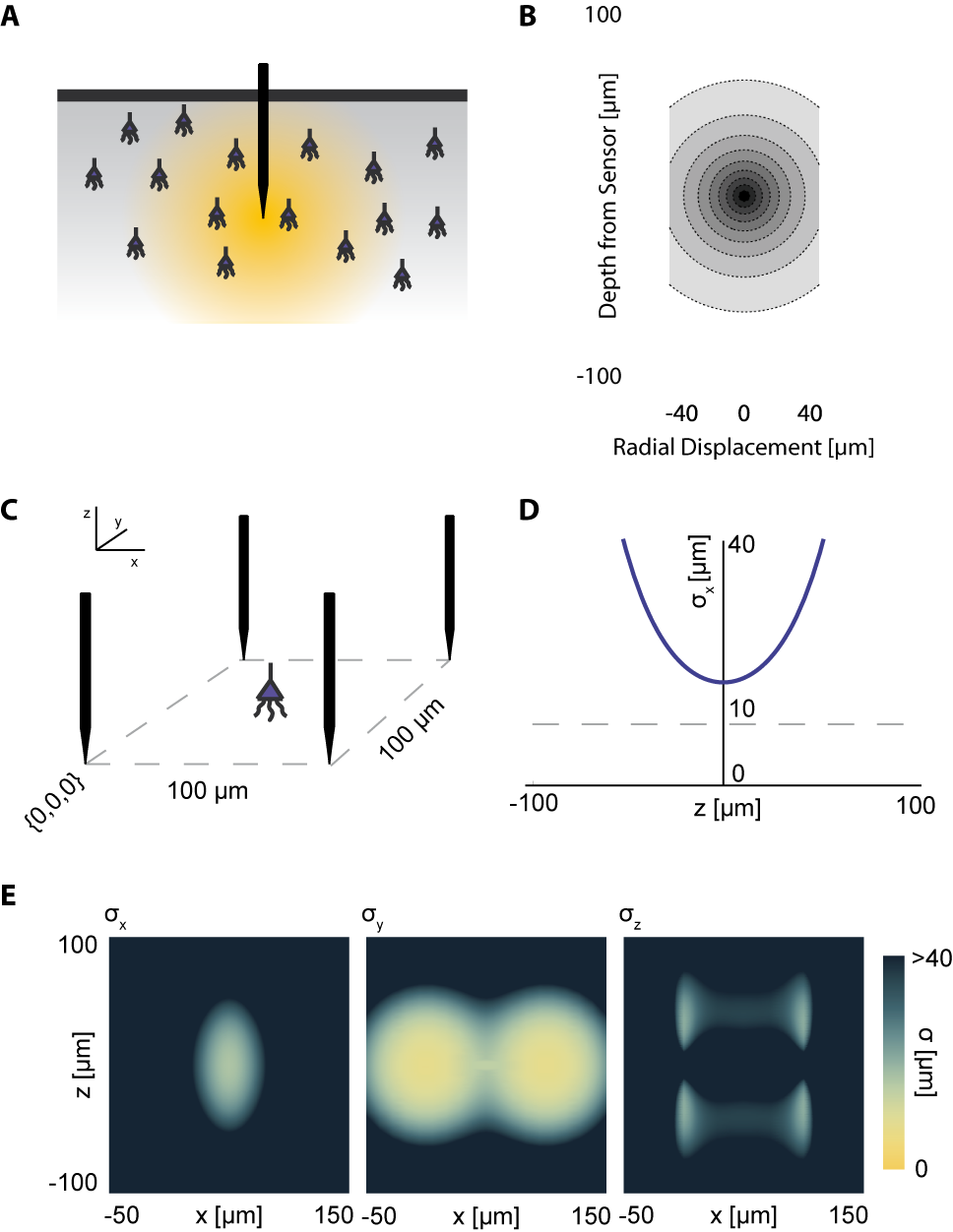}
\caption[Electrical Recording]{\textbf{Electrical Recording.} An overview of the modeling and Fisher information analysis of electrical recording. \textbf{(A)} Schematic: An electrode records electrical signals directly from nearby neurons. \textbf{(B)} The spatial PSF for a single electrode recording, valued in arbitrary units, for an electrode located at (0,0,0). \textbf{(C)} A schematic for the simple 4-electrode recording system simulated here. Electrodes are arranged in a $\SI{100}{\micro\meter} \times \SI{100}{\micro\meter}$ square, all with $z = 0$. The coordinate system for \textbf{(D)} and \textbf{(E)} is defined. \textbf{(D)} The standard deviation of an estimator for position on the $x$ axis ($\sigma_x$) for a source located at $(50,50,z)$. The grey dashed line indicates a CRB standard deviation of \SI{10}{\micro\meter}. This \SI{10}{\micro\meter} standard deviation corresponds to a 95\% accuracy of determining the correct active neuron for neurons whose centers are \SI{40}{\micro\meter} apart, and assuming a Gaussian estimation profile. \textbf{(E)} Standard deviation of estimators for $x$, $y$, and $z$ location ($\sigma_x$, $\sigma_y$, $\sigma_z$) for a source located at $(x,50,z)$. See \autoref{tab:fieqs} and \autoref{tab:2} for equations and parameters used to generate this figure.}
\label{fig:3}
\end{figure}

The electrical potential from an isolated firing neuron decays approximately exponentially with increasing distance~\cite{segev2004,gray1995}, at least at short distances. Here, we model a simple electrical system: an isotropic electrode with spherical symmetry. In this isotropic approximation, the PSF has an exponential decay with radial distance from the electrode tip (PSF taken from \autoref{tab:fieqs}, using parameters found in \autoref{tab:2}; \autoref{fig:3}B).

For electrical recording, estimators of location parameters have the lowest standard deviation $\sigma_x$ and $\sigma_y$ when in-between two electrodes, and the lowest $\sigma_z$ when directly above or below an electrode (\autoref{fig:3}D\&E). Generally, we see that electrical recordings provide relatively weak information over a relatively wide area. In fact, we find that, in ``worst-case'' regions, standard electrode arrays should have difficulty localizing a source within the bounds required to discriminate between neighboring neurons. Given that current arrays generally require more information than a single sample of signal intensity to sort spikes (e.g. waveform shape is used), this is an expected result. 

\subsection{Optical Sensing}
\subsubsection{General Information}
Optical recording of neural activity generally relies on fluorescent dyes that are sensitive to activity. In order to measure this signal, a neuron must be illuminated with light in the dye's excitation spectrum. Light is then emitted by the dye at a distinct, longer (lower energy) wavelength, which is picked up by a photodetector. Optical signal transmission is subject to absorption, scattering, and diffraction, which degrade the emitted signals with distance. Absorption of light effectively cause an exponential decrease in intensity of detected photons as light travels through a medium~\cite{lambert1892,theer2006}. Scattering can affect light in multiple ways; high-angle scattering diverts photons from the detector and produces an effect similar to absorption, while low-angle scattering causes blurring of the image on the detector. This blurring increases approximately linearly with depth into the tissue~\cite{tian2011}. Finally, diffraction results when light passes through an aperture, creating the finite-width Airy disk~\cite{airy1835}. In our optical PSFs, we assume scattering and diffraction result in Gaussian blurring~\cite{tian2011,thomann2002}. Our PSFs assume imaging through a single homogeneous medium; in practice, tissue inhomogeneity and refractive index mismatch can produce additional aberrations in the absorption, scattering, and diffraction domains that we do not model here.

In a typical optical setup, a lens focuses a set of photons from one point in space onto a corresponding point behind the lens. This phenomenon can be used either to focus incident light onto a desired location for illumination, or to focus emitted light from the “focal plane” onto a photodetector for imaging. Photons from outside the focal plane will be blurred, and this blurring increases linearly as distance from a focus point increases~\cite{kirshner2013,torreao2005}. We also assume defocusing results in Gaussian blurring~\cite{kirshner2013,torreao2005}.

\subsubsection{Wide-field Fluorescence Microscopy}
\begin{figure}[H]
\centering
\includegraphics[scale=0.5]{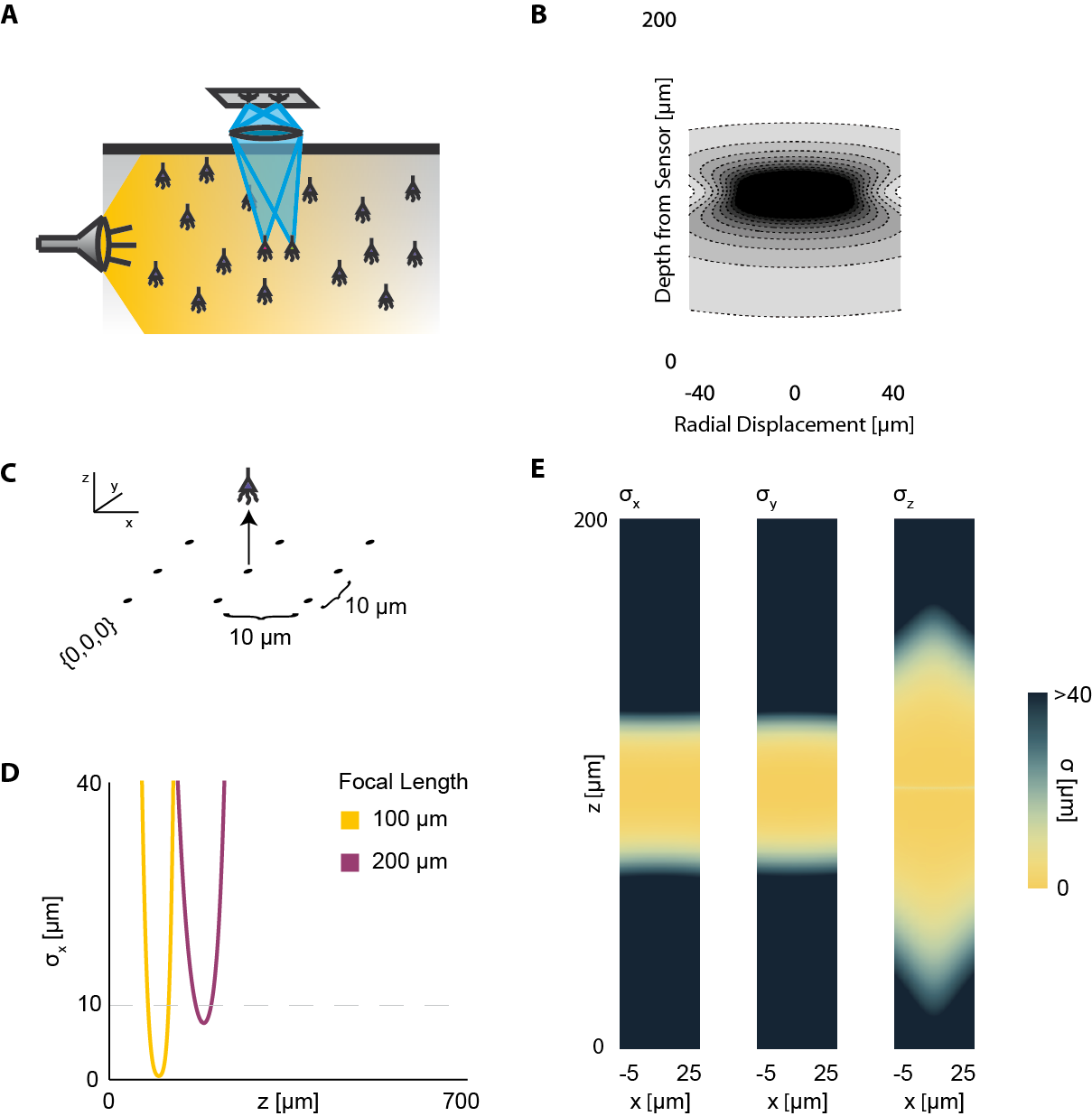}
\caption[Wide-field Fluorescence Optical Recording]{\textbf{Wide-field Fluorescence Optical Recording.} An overview of the modeling and Fisher information analysis of wide-field fluorescence optical recording. \textbf{(A)} Schematic: The whole recording volume is illuminated; dye in active neurons fluoresces and emits light; the emitted light is focused by a lens onto a photosensor. \textbf{(B)} The spatial PSF for wide-field fluorescence optical recording, valued in arbitrary units, for a lens centered at (0,0,0) with a focal plane at \SI{100}{\micro\meter}.  \textbf{(C)} A schematic for the simple 9-sensor optical recording system simulated here. Sensors are arranged in a $3 \times 3$ grid with a pitch of \SI{10}{\micro\meter}, all sensors with $z = 0$. The coordinate system for \textbf{(D)} and \textbf{(E)} is defined. \textbf{(D)} The standard deviation of an estimator for position on the $x$ axis ($\sigma_x$) for a source located at $(10,10,z)$ and an optical system with focal depth of either \SI{100}{\micro\meter} or \SI{200}{\micro\meter}. The grey dashed line indicates a CRB standard deviation of \SI{10}{\micro\meter}. \textbf{(E)} Standard deviation of estimators for $x$, $y$, and $z$ location ($\sigma_x$, $\sigma_y$, $\sigma_z$) for a source located at $(x,10,z)$ and an optical system with focal depth of \SI{100}{\micro\meter}. See \autoref{tab:fieqs} and \autoref{tab:2} for equations and parameters used to generate this figure.}
\label{fig:4}
\end{figure}

Neural activity in a focused optical system is generally sensed using fluorescent dyes, which require some excitatory light. In the canonical optical example of wide-field microscopy, an entire volume is illuminated (\autoref{fig:4}A). The PSF for this technology takes the above effects of absorption, scattering, diffraction, and defocusing into account; we assume total illumination so that the PSF here models the spread of the emission light (\autoref{fig:4}B, PSF taken from \autoref{tab:fieqs} using parameters found in \autoref{tab:2}).

For optical recording with a simple lens, estimators of location parameters have lowest standard deviation $\sigma_x$, $\sigma_y$, and $\sigma_z$ when centered above the imaging system in the focal plane (\autoref{fig:4}D\&E). For large depth, the ability to distinguish locations decreases rapidly due to photon loss caused by scattering and absorption (\autoref{fig:4}D\&E). For medium depth ranges, scattering blurs the image, even on the focal plane. These phenomena decrease the utility of deep focal-plane wide-field optics in tissue. At shallower focal depths, optical recordings provide a large amount of information on the focal plane, while carrying relatively little information about sources out of the focal plane (\autoref{fig:4}D\&E).

\subsubsection{Two-photon Microscopy}
\begin{figure}[H]
\centering
\includegraphics[scale=0.5]{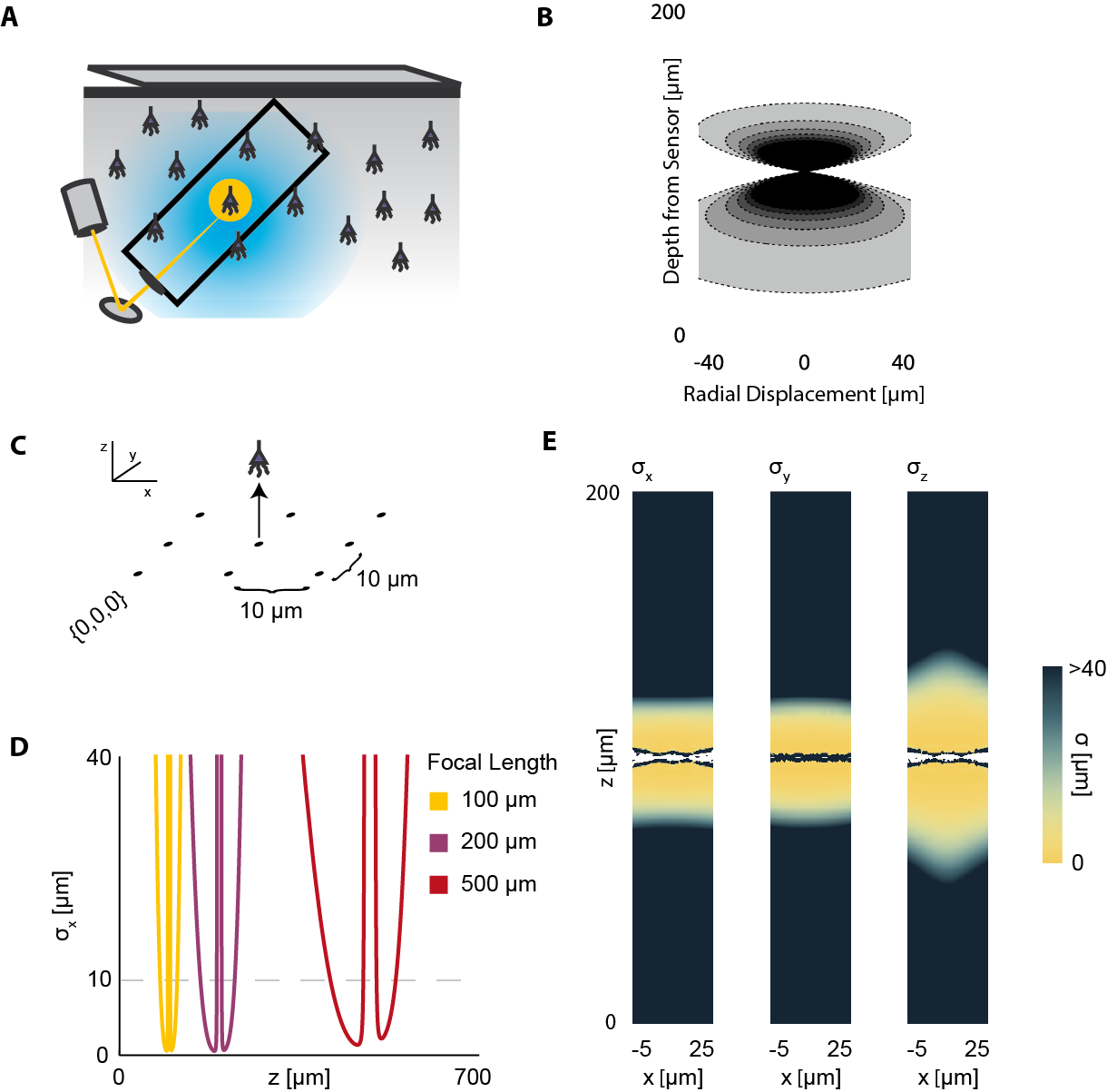}
\caption[Two-photon Optical recording]{\textbf{Two-photon Optical Recording.} An overview of the modeling and Fisher information analysis of 2-photon optical recording. \textbf{(A)} Schematic: incident light is focused onto a particular location in a volume; dye in neurons illuminated by the incident light fluoresces and emits light; the emitted light is sensed by a large single photosensor. The black box indicates the space represented in \textbf{B}, with zero depth being located at the lens and increasing depth indicating increasing distance into the brain. \textbf{(B)} The spatial PSF for incident light relative to its source in 2-photon optical recording. It is valued in arbitrary units for a lens centered at (0,0) with a focal plane at \SI{100}{\micro\meter}. \textbf{(C)} A schematic for the simple 9-pixel two-photon recording system simulated here. Sampled points are arranged in a $3 \times 3$ grid with a pitch of \SI{10}{\micro\meter}, all points with $z = 0$. The coordinate system for \textbf{(D)} and \textbf{(E)} is defined. \textbf{(D)} The standard deviation of an estimator for position on the $x$ axis ($\sigma_x$) for a source located at $(10,10,z)$ and an optical system with focal depth of \SI{100}{\micro\meter}, \SI{200}{\micro\meter}, or \SI{500}{\micro\meter}. The grey dashed line indicates a CRB standard deviation of \SI{10}{\micro\meter}. \textbf{(E)} Standard deviation of estimators for $x$, $y$, and $z$ location ($\sigma_x$, $\sigma_y$, $\sigma_z$) for a source located at $(x,10,z)$ and an optical system with focal depth of \SI{100}{\micro\meter}. White regions indicate regions where the Fisher information matrix is ill-conditioned. See \autoref{tab:fieqs} and \autoref{tab:2} for equations and parameters used to generate this figure.}
\label{fig:5}
\end{figure}

In two-photon microscopy, long-wavelength incident light (i.e. composed of low-energy photons) is focused onto a single point of interest to excite fluorophores in that area. In order for the fluorophore to emit light, two low-energy photons must be absorbed nearly simultaneously; the likelihood of this event is proportional to the square of the intensity of incident light at a point. Effectively, this concentrates the area of sufficient illumination to a volume nearby the focal point of the incident beam (while increasing the illumination power requirements)~\cite{helmchen2005}. Like with wide-field fluorescence microscopy, the PSF is a function of defocusing, absorption, and scattering (\autoref{fig:5}B, PSF taken from \autoref{tab:fieqs} using parameters found in \autoref{tab:2}). We assume total photon capture so that the PSF here models the spread of the excitation light.

For two-photon microscopy, estimators of location parameters have lowest standard deviation $\sigma_x$, $\sigma_y$, and $\sigma_z$ just above and below the focal plane (\autoref{fig:5}C). Perhaps counter-intuitively, there are extremely-high or undefined $\sigma$'s along the focal plane. This is due to our simplified recording setup: given the tightly-focused PSF for two-photon microscopy, sources very close to the focal plane of our setup are effectively only ``seen'' by one sensor. Thus, we cannot gather meaningful information about the source's three location parameters, resulting in a singular or near-singular Fisher information matrix. In practice, this is alleviated by either decreasing the pitch of sensed regions or applying magnification to the sample, which we do not model here. We also see a reduced dependence on focal depth when compared to a wide-field imaging setting, as expected.

\begin{table}[H]
\centering
\begin{tabular}{| p{2.25cm} | l |}
\hline
\textbf{Electrical} & 
\begin{tabular}[l]{l}
$w_{el}(r) = \exp \left(\frac{ - r}{C_{el}} \right) $
\end{tabular} \\
 \hline

\textbf{Optical: Wide-field fluorescence microscopy}  &
\begin{tabular}[l]{l}
$w_{wf}(\ell,z) = \frac{Q}{{2\pi }}\exp \left( {\frac{ - z}{C_{op}}} \right)\frac{1}{{\left( {s_{defocus}^2 + s_{dif}^2 + s_{scat}^2} \right)}}$ \\
$\quad \quad \times \exp \left( {\frac{{ - {\ell ^2}}}{{2\left( {s_{defocus}^2 + s_{dif}^2 + s_{scat}^2} \right)}}} \right)$ \\[0.5cm]

$s_{defocus}=\frac{D_{lens}\cdot(z_0-z)}{2z_0}$, $s_{dif} = \frac{0.42\lambda\cdot z}{D_{lens}}$, $s_{scat} = \gamma z$
\end{tabular} \\
 \hline

\textbf{Optical: 2-photon microscopy}  &
\begin{tabular}[l]{l}
$w_{2P}(\ell,z) = \frac{1}{\pi }\frac{1}{{\left( {s_{defocus}^2 + s_{dif}^2 + s_{scat}^2} \right)}}$\\
$\quad \quad \times{\left( {Q\exp \left( {\frac{{ - z}}{{{C_{op}}}}} \right)\exp \left( {\frac{{ - {\ell ^2}}}{{2\left( {s_{defocus}^2 + s_{dif}^2 + s_{scat}^2} \right)}}} \right)} \right)^2}$
\end{tabular} \\
\hline
\end{tabular}

\caption[Point Spread Functions of Recording Modalities]{\textbf{Point Spread Functions of Recording Modalities.} Analytic expressions are given for PSFs. $r$ is the distance in any radial direction from the electrode, and $\ell$ is the lateral distance from the center of the lens for optical techniques. Note that $r^2=x^2+y^2+z^2$ and $\ell^2=x^2+y^2$. $C_{el}$ is the spatial constant of electrical decay. $C_{op}$ is the spatial constant of optical decay. $s^2_{defocus}$, $s^2_{dif}$, and $s^2_{scat}$ are the variance of the spread of optical light due to defocusing, diffraction, and scattering, respectively. $D_{lens}$ is the diameter of a lens. $\lambda$ is the wavelength of the light. $z_0$ is the focus depth, and $Q$ is the light flux (area per photon).}
\label{tab:fieqs}
\end{table}

\begin{table}
\centering
\begin{tabular}{| l | l |}
\hline
\textbf{Parameter} & \textbf{Value} \\ \hline
$C_{el}$ & \SI{28}{\micro\meter}~\cite{segev2004,gray1995} \\ \hline
$D_{lens}$ & \SI{300}{\micro\meter} (within current dimensions) \\ \hline
$\lambda$ (wide-field) & \SI{633}{\nano\meter} (visible light) \\ \hline
$\lambda$ (2-photon) & \SI{800}{\nano\meter} \\ \hline
$\gamma$ (wide-field) & \num{0.15}~\cite{tian2011,orbach1983} \\ \hline
$C_{op}$ (wide-field) & \SI{100}{\micro\meter} (with \SI{515}{\nano\meter} light)~\cite{theer2006} \\ \hline
$\gamma$ (2-photon) & \num{0.002} (with \SI{725}{\nano\meter} light)~\cite{chaigneau2011} \\ \hline
$C_{op}$ (2-photon) & \SI{200}{\micro\meter} (with \SI{909}{\nano\meter} light)~\cite{theer2006}\\ \hline
\end{tabular}
\caption[Simulation Parameter Values]{\textbf{Simulation Parameter Values}}
\label{tab:2}
\end{table}

\subsection{Technological Optimization}

This example will demonstrate the ability to use Fisher information to ask questions about the necessary experimental parameters of neural recording technologies. In particular, we will use Fisher information to examine sensor placement in electrical recording. In order to successfully record activity from every neuron in a volume, we must place sensors so that they extract sufficient information about every neural location in that volume. That is, the CRB regarding the ability to estimate the location of each point in a volume must be below some threshold for localization. 

\begin{figure}[H]
\centering
\includegraphics[scale=0.6]{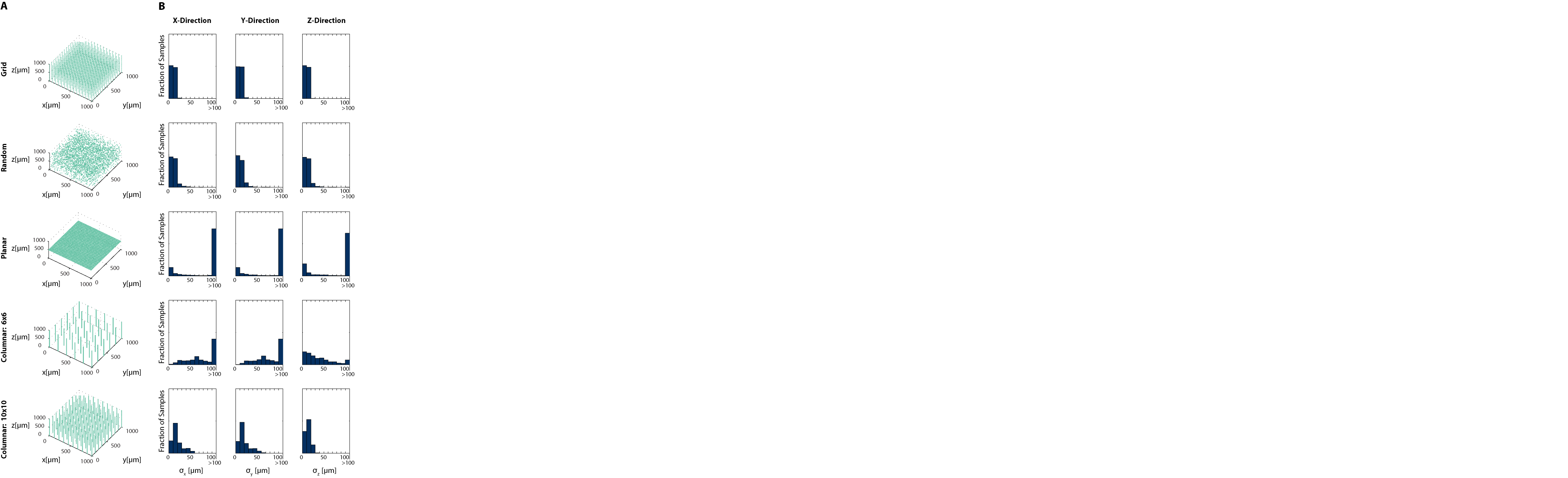}
\caption[Electrode Placement and Fisher Information]{\textbf{Electrode Placement and Fisher Information.} CRBs on the x, y, and z coordinates of neurons using various electrode arrays. We simulate $\sim\num{3.5e3}$ electrodes in a $\SI{1}{\milli\meter}\times\SI{1}{\milli\meter}\times\SI{1}{\milli\meter}$ cube of brain tissue. Electrodes were arranged in one of five patterns: uniformly distributed in a grid throughout the volume (top row), random placement (second row), electrodes uniformly distributed on a plane at \SI{500}{\micro\meter} depth (third row), a $6\times6$ grid of columns of electrodes with 100 electrodes evenly distributed in each column (fourth row), and a $10\times10$ grid of columns of electrodes with 30 electrodes evenly distributed in each column (bottom row). Total Fisher information about a point consists of the sum of information contained about that point in each sensor. \textbf{(A)} Distribution of electrodes in the volume for each pattern. \textbf{(B)} Distribution of Cramer-Rao bounds about a random sample of $10^4$ points in the volume. Standard distributions are shown. The three columns represent estimation about the x, y, and z coordinates, from left to right. See \autoref{tab:2} for parameter values.}
\label{fig:6new}
\end{figure}

Here, we simulate several possible arrangements of electrical sensors and evaluate the information that these systems provide about different locations in a volume. Specifically, we look at five electrode arrangements: (1) electrodes evenly distributed in an equilateral grid (Grid electrodes); (2) randomly placed electrodes (Random electrodes); (3) electrodes evenly distributed in a plane (Planar electrodes); and (4\&5) two arrangements of columns of electrodes, where electrodes are densely packed within a column, and these columns are arranged in a grid~\cite{zorzos2012} (Column electrodes) (\autoref{fig:6new}A). Here, we assume that noise is independent between sensors, i.e. noise is all on the sensor. Under this assumption, each electrode takes an independent sample of a signal; information about the location of the source of that signal is then additive across sensors. Fisher information here is thus the information the entire ensemble of electrodes provides about a point. In this simplified example, we determine localization, rather than resolution, capabilities, which corresponds to the common situation of sparse neural firing. Multiple sources would necessarily reduce the amount of information contained about individual sources and would be geometry-dependent.

In this simplified simulation, Grid electrodes and Random electrodes have the best performance, as they sample space uniformly (Grid) or almost uniformly (Random) (\autoref{fig:6new}B). Due to the regular nature of Grid electrodes, there is the added benefit of a guaranteed lower bound for information carried about locations in a volume. Planar electrodes are able to estimate a small fraction of locations very well, but  carry very little information about most locations in a volume. Columnar electrodes, in general, have the interesting property that the z coordinate can be estimated more accurately, due to the density of electrodes in this direction. It's also important to note that the feasibility of Columnar electrodes will likely depend on the spacing between shanks. As the shanks move closer together (e.g. the bottom row compared to the fourth row), a greater number of neurons will able to be distinguished. The use of this Fisher information framework promises to inform sensor placement decisions.

\subsection{Technology Comparison}
Finally, we demonstrate the use of Fisher information for determining resolution ability. This example will demonstrate the ability to use Fisher information to compare technologies. In order to determine appropriate technologies for a given situation, it is necessary to know which technology will maximize the information output, and where information will be concentrated for a given technology. 

\begin{figure}
\centering
\includegraphics[scale=0.5]{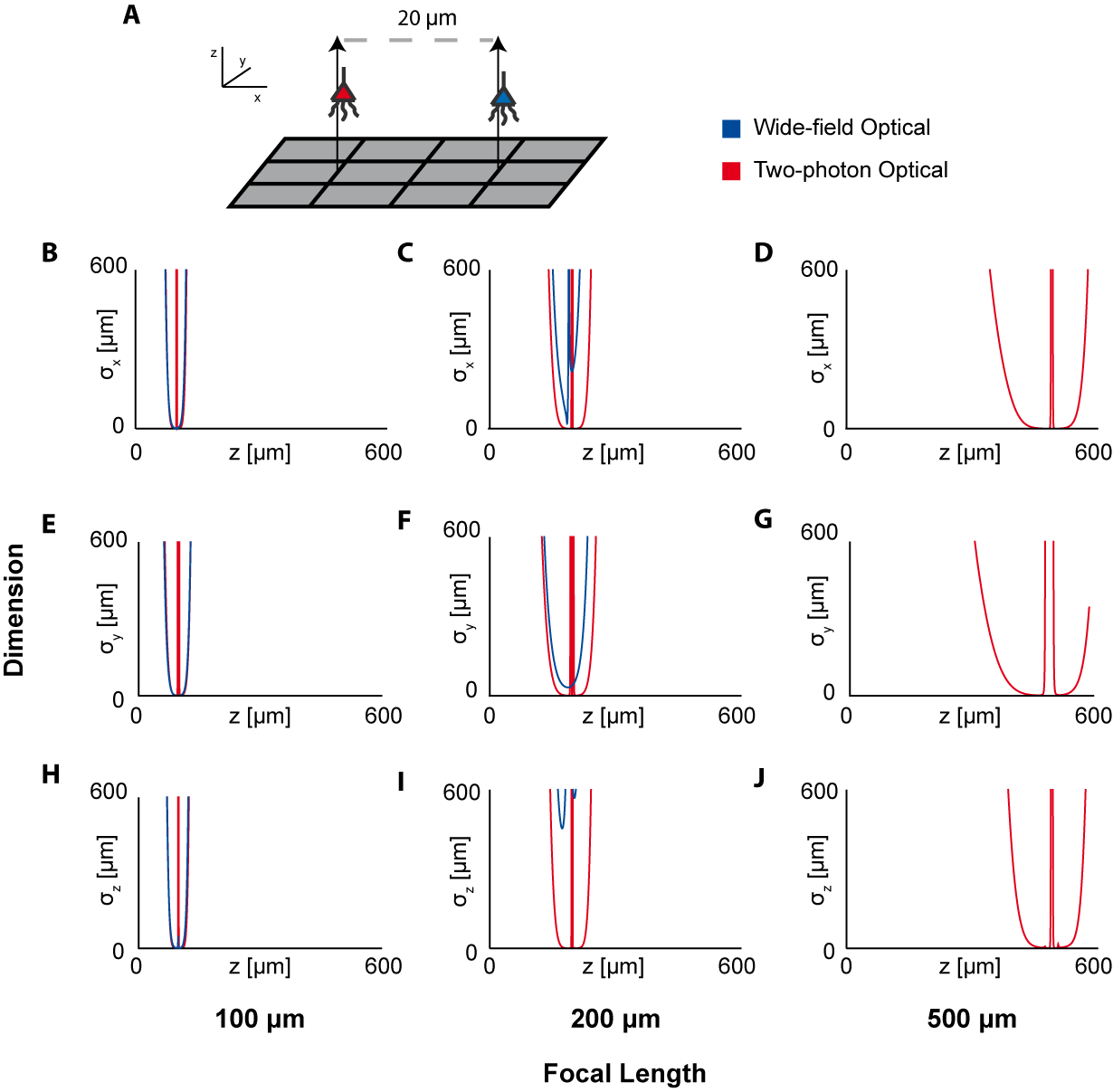}
\caption[Optical Technology Comparison at Multiple Focal Depths]{\textbf{Optical Technology Comparison at Multiple Focal Depths.} CRB on the location of the x, y, and z coordinates of a source in a multi-sensor, two-source system. The depth of the sources is varied by an equal amount and the CRB on each of the sources is calculated at each depth (the CRBs of only one source is shown; they are equivalent due to the symmetric setup). This analysis is performed for wide-field fluorescence and two-photon optical systems. \textbf{(A)} Schematic of recording system: An evenly-spaced $4 \times 3$ grid of sensors detects two sources. Sensed regions have pitch of \SI{10}{\micro\meter}, and neurons are separated on the $x$-axis by \SI{20}{\micro\meter}. \textbf{(B,E,H)} CRBs with a focal depth of \SI{100}{\micro\meter}. \textbf{(C,F,I)} CRBs with a focal depth of \SI{200}{\micro\meter}. \textbf{(D,G,J)} CRBs with a focal depth of \SI{500}{\micro\meter}. CRBs for the x, y, and z coordinates are in the first, second, and third rows, respectively, and are reported as standard deviations. See \autoref{tab:2} for parameter values. }
\label{fig:8new}
\end{figure}

Here we apply this Fisher information framework to a two-source, multi-sensor setup for both wide-field fluorescence and two-photon microscopy in order to determine performance over depth (\autoref{fig:8new}). We find, perhaps confirming intuition, that wide-field and two-photon fluorescence perform similarly for shallow sections, but performance of wide-field fluorescence microscopy degrades significantly at a depth of \SI{500}{\micro\meter} while two-photon performs well at this depth. Interestingly, both methods contain a large amount of information not only about signals near the focal point, but also about sources nearby the lens. This implies that signals could be recovered from out-of-focus samples given proper recording conditions. While this demonstration yielded the expected results, this framework could be used to compare existing technologies in novel situations, or to compare novel technologies.

\section{Demonstrations Discussion}
\label{sec:discussion}
We have demonstrated how the Fisher information framework can be applied to neural recording technologies, and have demonstrated possible applications of this framework including determining optimal technology design and comparing technologies under differing recording conditions. In these demonstrations, interesting findings emerged, some of which confirm experimental knowledge. For instance, (1) when using columnar electrodes, increasing the spacing between electrode shanks leads to a very large fall-off in the number of neurons that can be recorded. (2) For shallow recording depths, wide-field and two-photon microscopy have similar performance capabilities, but at larger depths two-photon microscopy becomes significantly better. 
 
We made several simplifications regarding neural activity, noise, and recording technologies when demonstrating the use of the Fisher information framework. However, these approximations were useful in demonstrating a unifying view over recording methodologies in a single paper. Moreover, much is still experimentally unknown about noise sources and their relation to neural activity. While our demonstrations cannot give precise predictions about the capabilities of recording technologies, they demonstrate general scaling properties of the technologies, as well as illustrate situations in which the framework could be useful with more detailed models of neural recording.
 
A first simplification is that our demonstrations used approximate models of how neurons and noise affect sensor signals. Our demonstrations showed how we could use recording channels to identify the location of a fixed, known, activity. In practice these activities fluctuate over time, and can differ based on the type of neuron. In addition, we assumed that the effects of neural activity are linearly combined into the sensor signal. In practice, nonlinear effects such as sensor saturation may be important. Both can be incorporated into a Fisher information-based framework, although neither are treated here. Perhaps the largest simplifying model, the various noise sources were approximated by a simple function that ignores many potential sources of noise (see \textit{\nameref{sec:supinfo}}). A comprehensive model of how noise affects neurons and sensors does not yet exist. Further research in this area will yield more informative results.
 
Second, we asked how we could use simplified models of recording systems to estimate the locations of neurons.  For example, for optical recordings we assumed scattering through homogenous tissue, and for electrical recordings we ignored the filtering properties of electrodes. There exists a rich literature of modeling optical and electrical systems that could allow better models of recording modalities (e.g.~\cite{theer2006,camunas2013}); incorporating these models into the framework may alleviate some of the concerns over oversimplification, and may even provide a framework for validating those models.

In order to calculate the Fisher information contained by a given technique, we need to know its PSF and noise sources. When a technology is developed, experimentally determining these functions would allow this Fisher information to accurately be applied. These Fisher information calculations could determine how optimal a technique's performance is. This information may then influence further design choices.

\section{Additional Methods}
\subsection{Noise Calculations}
In our Applications simulations, we make several assumptions about noise. We assume noise sources are uncorrelated (i.e. the noise from each neuron is independent and independently distributed). The sensor signal variance arises from signal dependent noise, with a standard deviation proportional to the mean signal. The signal dependent noise can be on all background neurons and/or on the sensor. As the mean activity is $I_0$, the standard deviation of the activity is $\alpha\cdot I_0$, where $\alpha$ is a constant. The activity that reaches the sensor $i$ (the signal) from a given neuron $j$ then has a variance of $\sigma _j^2 = \alpha \cdot {\left( {{I_0} \cdot w({\bs{d}^{i,j}})} \right)^2}$. As the noise sources are independent, their variances can be added, so $\sigma _{noise}^2 = \sum\limits_{j \ne k} {\sigma _j^2} $ (recall that we do not include noise from the neuron of interest). In simulations with two neurons of interest, we do not include noise from both neurons. We assume that neurons are uniformly distributed across the brain with density $\rho_{space}$ and that all neurons have the same probability of firing at a given time, $\rho_{fire}$.

\begin{equation}
\begin{split}
{\sigma^2 _{noise}} & = {\alpha _{sens}}{\rho _{fire}}{\rho _{space}}\int\limits_V {I_0^2{w^2}} dV + {\alpha _{neur}}{\rho _{fire}}{\rho _{space}}\int\limits_V {I_0^2{w^2}} dV\\
& = \alpha {\rho _{fire}}{\rho _{space}}\int\limits_V {I_0^2{w^2}} dV
\end{split}
\end{equation}
In our simulations, we set $\alpha=0.1$ (action potentials have SNRs ranging from 5-25~\cite{erickson2008}), $\rho_{fire}=0.01$ (assuming neurons on average fire at \SI{5}{\hertz}~\cite{harris2012} and action potentials last $\approx\SI{2}{\milli\second}$), and $\rho_{space}=\SI{67000}{\per\milli\meter\cubed}$ (dividing the number of neurons in the human brain, $\approx\num[exponent-product = \times]{8e10}$~\cite{azevedo2009} by its volume, $\approx\SI{1200}{\centi\meter\cubed}$~\cite{allen2002}).

\subsection{Applications: Electrode Grid Analysis}
Electrode locations were assigned to nodes on a \SI{1}{\micro\meter} grid spanning a $\SI{1}{\milli\meter}\times\SI{1}{\milli\meter}\times\SI{1}{\milli\meter}$ cube using the following procedures:

\textit{Columnar $6 \times 6$}: Column locations were spaced evenly, \SI{200}{\micro\meter} apart, on a $6 \times 6$ grid in the x-y plane. $101$ electrodes were distributed evenly along each column, \SI{10}{\micro\meter} apart.

\textit{Columnar $10 \times 10$}: Column locations were spaced evenly, \SI{111}{\micro\meter} apart, on a $10 \times 10$ grid in the x-y plane. $31$ electrodes were distributed evenly along each column, \SI{33}{\micro\meter} apart.

\textit{Random}: Locations on the grid were drawn from a uniform random distribution with replacement.

\textit{Planar}: Electrodes were placed on a uniform $61 \times 61$ grid in the x-y plane, corresponding to a grid spacing of \SI{17}{\micro\meter}, with a depth of \SI{500}{\micro\meter}.

\textit{Grid}: Electrodes were placed on a uniform $15 \times 15 \times 15$ grid in the volume, corresponding to a grid spacing of \SI{71}{\micro\meter}.

These procedures give locations for $3636$, $3100$, $3636$, $3721$, and $3375$ electrodes respectively.

For each arrangement of electrodes, we determine the $3 \times 3$ Fisher information matrix about the Cartesian location of a source contained by each electrode. We then sum these matrices under the assumption of independent noise on sensors, giving the aggregate Fisher information matrix for sensor $i$:

\begin{equation}
{\cal I}_{mn} = \frac{I_k^2}{\sigma _{noise}^2}  \sum\limits_{t = 1}^e \frac{\partial w({\bs{d}^{i,k}})}{\partial \theta_m} \frac{\partial w({\bs{d}^{i,k}})}{\partial \theta_n}
\end{equation}
Where ${I_k}$ is the intensity of the neuron of interest, $\theta_m, \theta_n \in \{x,y,z\}$ of the given neuron are the parameters of interest, and $e$ is the number of electrodes (See \textit{\nameref{sec:supinfo}}). The Cramer-Rao bounds on $x,y,z$  are the diagonal elements of $\mathcal{I}^{-1}$.

\section{Acknowledgements}
We would like to thank Dario Amodei and Darcy Peterka for their helpful comments. We would like to thank Dan Dombeck for helpful discussions regarding optics and Mikhail Shapiro for discussions regarding MR applications.

Thaddeus Cybulski, Joshua Glaser, and Bradley Zamft are supported by NIH grant 5R01MH103910. Adam Marblestone is supported by the Fannie and John Hertz Foundation fellowship. Konrad Kording is funded in part by the Chicago Biomedical Consortium with support from the Searle Funds at The Chicago Community Trust. Konrad Kording is also supported by NIH grants 5R01NS063399, P01NS044393, and 1R01NS074044. George Church acknowledges support from the Office of Naval Research and the NIH Centers of Excellence in Genomic Science. Edward Boyden acknowledges funding by Allen Institute for Brain Science; AT\&T; Google; IET A.~F.~Harvey Prize; MIT McGovern Institute and McGovern Institute Neurotechnology (MINT) Program; MIT Media Lab and Media Lab Consortia; New York Stem Cell Foundation-Robertson Investigator Award; NIH Director's Pioneer Award 1DP1NS087724, NIH Transformative Awards 1R01MH103910 and 1R01GM104948, NSF INSPIRE Award CBET 1344219, Paul Allen Distinguished Investigator in Neuroscience Award; Skolkovo Institute of Science and Technology; Synthetic Intelligence Project (\& its generous donors).

\bibliographystyle{unsrt}
\bibliography{InfoBAMlib}

\section{Supplementary Information}
\label{sec:supinfo}
\subsection{Noise Sources}
The Fisher information framework allows for arbitrary noise sources, so long as they are able to be modeled. However, to demonstrate potential applications, we used a very simplified noise model that only considered signal dependent noise where the standard deviation was proportional to the mean.

There are multiple potentially relevant sources of noise that could readily be included in our model. (1) Each sensor has a constant level of noise simply due to thermal effects. (2) Many sensors have an additional variance that is proportional to the square of the signals, e.g. reference fluctuations. (3) Many sensors have an additional variance that is proportional to the signal, e.g. due to low numbers of photons (shot noise). (4,5) Each neuron may produce constant noise, e.g. background fluorescence of dyes. These neural noise sources may be independent or correlated. (6,7) Each neuron may produce variance that quadratically depends on its activation, e.g. action potentials that propagate back into varying parts of the dendritic tree.\footnote{In a simplistic model, when a neuron fires, the action potential spreads into some variable proportion of the dendritic tree. If the recorded signal is dependent on the proportion of dendritic branches the action potential propagates into, then the standard deviation of the recorded signal is proportionate to the mean signal entering the dendrites.} These neural noise sources may be independent or correlated. (8,9) Each neuron may produce variance that linearly depends on its signal strength, e.g. fluorophore activations. These neural noise sources may be independent or correlated. We have some knowledge about the exact sizes of these signals~\cite{marblestone2013}, but most of these numbers are hard to know. They may be reasonable to measure in future experiments.

Taking these signals together, we obtain the following noise level on a sensor $i$ (given a recording of $N$ firing neurons indexed by $j$):

\begin{equation}
\begin{split}
{\sigma ^2}_{noise} =& \underbrace {\sigma _{sens}^2}_{constant\,sensor\,noise} + \underbrace {{\alpha _1}\sum\limits_{j = 1}^N {{I_0}^2{\left(w(\bs{d}^{i,j})\right)}^2} }_{ind\,sensor\,\sigma \propto \mu \,SDN} + \underbrace {{\alpha _2}\sum\limits_{j = 1}^N {{I_0}{\left(w(\bs{d}^{i,j})\right)}^2} }_{ind\,sensor\, {\sigma ^2} \propto \mu\,SDN}\\ 
&+ \underbrace {{\alpha _3}N \cdot \sigma _{neur}^2}_{constant\,ind\,neuron\,noise}+ \underbrace {{\alpha _4}N \cdot \sigma _{neur}^2}_{constant\,corr\,neuron\,noise}\\
&+ \underbrace {{\alpha _5}\sum\limits_{j = 1}^N {{I_0}^2{\left(w(\bs{d}^{i,j})\right)}^2} }_{ind\,neuron\,\sigma \propto \mu \,SDN} + {\underbrace {{\alpha _6}\left( {\sum\limits_{j = 1}^N {{I_0}{\left(w(\bs{d}^{i,j})\right)}} } \right)^2}_{corr\,neuron\,\sigma \propto \mu \,SDN}} \\
&+ \underbrace {{\alpha _7}\sum\limits_{j = 1}^N {{I_0}{\left(w(\bs{d}^{i,j})\right)}} }_{ind\,neuron\,{\sigma ^2} \propto \mu\,SDN} + {\underbrace {{\alpha _8}\left( {\sum\limits_{j = 1}^N {\sqrt {{I_0}{\left(w(\bs{d}^{i,j})\right)}} } } \right)^2}_{corr\,neuron\,{\sigma ^2} \propto \mu\,SDN}}
\end{split}
\end{equation}
where $ind$ and $corr$ refer to independent and correlated noise sources, and $SDN$ refers to signal dependent noise.

Assuming, as we do in the main text's demonstrations, that neurons are uniformly distributed and have a uniform firing rate across the entire volume:

\begin{equation}
\begin{split}
{\sigma ^2}_{noise} =& \underbrace {\sigma _{sens}^2}_{constant\,sensor\,noise} + \underbrace {{\alpha _1}{\rho _{fire}}{\rho _{space}}\int {{I_0}^2{w^2}} dV}_{ind\,sensor\,\sigma \propto \mu \,SDN} + \underbrace {{\alpha _2}{\rho _{fire}}{\rho _{space}}\int {{I_0}w} dV}_{ind\,sensor\,{\sigma ^2} \propto \mu\,SDN}\\
&+ \underbrace {{\alpha _3}{\rho _{fire}}{\rho _{space}}V \cdot \sigma _{neur}^2}_{constant\,ind\,neuron\,noise} + \underbrace {{\alpha _4}{\rho _{fire}}{\rho _{space}}V \cdot \sigma _{neur}^2}_{constant\,corr\,neuron\,noise}\\
& + \underbrace {{\alpha _5}{\rho _{fire}}{\rho _{space}}\int {{I_0}^2{w^2}} dV}_{ind\,neuron\,\sigma \propto \mu \,SDN} + {\underbrace {{\alpha _6}{\rho _{fire}}{\rho _{space}}\left( {\int {{I_0}w} dV} \right)^2}_{corr\,neuron\,\sigma \propto \mu \,SDN}}\\
&  + \underbrace {{\alpha _7}{\rho _{fire}}{\rho _{space}}\int {{I_0}w} dV}_{ind\,neuron\,{\sigma ^2} \propto \mu\,SDN} + {\underbrace {{\alpha _8}{\rho _{fire}}{\rho _{space}}\left( {\int {\sqrt {{I_0}w} } dV} \right)^2}_{corr\,neuron\,{\sigma ^2} \propto \mu\,SDN}}
\end{split}
\end{equation}
Both constant and shot noise terms can be minimized in their effect by optimizing the experimental design, e.g. through good dyes and strong illumination (but see~\cite{marblestone2013}).

In addition, in the main text we assume that the noise is Gaussian, which has also been assumed in previous statistical formulations~\cite{shahram2005,shahram2006}. This assumption has been shown to be valid for thermal noise and shot noise in some conditions~\cite{tyson1999,bovik2010}.

\subsection{Fisher Information Derivation}

We have a distribution ${f}(X_i;\bs{\theta}) = {\mathcal N}\left( {\sum\limits_j {{I_j}w({\bs{d}^{i,j}})} ,\sigma _{noise}^2} \right)$ (See \eqref{eq:normtog}). Focusing on a particular neuron with index $k$, this equation becomes
\begin{equation}
{f_i}(X;\bs{d}) = {\cal N}\left( {\left( {{I_k}w({\bs{d}^{i,k}}) + \sum\limits_{j \ne k} {{I_j}w({\bs{d}^{i,j}})} } \right),\sigma _{noise}^2} \right).
\end{equation}
We are interested in finding its Fisher information about the parameter vector $\bs{\theta}$ that contains the directions of interest. Let $\theta_m$ be $d_p^{i,k}$ and $\theta_n$ be $d_q^{i,k}$, where $d_p^{i,k}$ and $d_q^{i,k}$ are two different location parameters of neuron $k$ with respect to sensor $i$. For simplicity of notation, we let $B = \sum\limits_{j \ne k} {{I_j}w({\bs{d}^{i,j}})} $.

\begin{equation}
\begin{aligned}
\finf_{mn} =& E \left[ \left( \frac{\partial}{\partial \theta_m} ln \left( f \left( X ; \bs{\theta} \right) \right) \right) \left( \frac{\partial}{\partial \theta_n} ln \left( f \left( X ; \bs{\theta} \right) \right) \right)\right]\\
=& E \left[ \left( \frac{1}{f \left( X ; \bs{\theta} \right)}  \frac{\partial}{\partial \theta_m} f \left( X ; \bs{\theta} \right)  \right) \left( \frac{1}{f \left( X ; \bs{\theta} \right)}  \frac{\partial}{\partial \theta_n} f \left( X ; \bs{\theta} \right)  \right) \right]\\
\end{aligned}
\end{equation}

\begin{equation}
\begin{aligned}
\frac{\partial}{\partial \theta_m} ln \left( f \left( X ; \bs{\theta} \right) \right) =& \frac{1}{f \left( X ; \bs{\theta} \right)} \frac{\partial}{\partial \theta_m} f \left( X ; \bs{\theta} \right)\\
=& \frac{1}{\frac{1}{\sqrt{2\pi\sigma_{noise}^2}} \exp \left( {\frac{{ - {{\left( {X - \left( {{I_k}w({\bs{d}^{i,k}}) + B} \right)} \right)}^2}}}{{2\sigma _{noise}^2}}} \right)}\cdot \\
&\frac{1}{\sqrt{2\pi\sigma_{noise}^2}} \exp \left( {\frac{{ - {{\left( {X - \left( {{I_k}w({\bs{d}^{i,k}}) + B} \right)} \right)}^2}}}{{2\sigma _{noise}^2}}} \right)\cdot \\
&\frac{\partial}{\partial \theta_m}\left( {\frac{{ - {{\left( {X - \left( {{I_k}w({\bs{d}^{i,k}}) + B} \right)} \right)}^2}}}{{2\sigma _{noise}^2}}} \right)\\
=& \frac{I_k}{2\sigma _{noise}^2} 2\left( {X - \left( {{I_k}w({\bs{d}^{i,k}}) + B} \right)} \right) \frac{\partial w({\bs{d}^{i,k}})}{\partial \theta_m}\\
=& \frac{I_k}{\sigma _{noise}^2} \frac{\partial w({\bs{d}^{i,k}})}{\partial \theta_m} \left( X - \left( I_k w(\bs{d}^{i,k}) + B \right) \right)
\end{aligned}
\end{equation}

\begin{equation}
\begin{aligned}
\finf_{mn} =& E \left[ \left( \frac{I_k^2}{\sigma _{noise}^4} \frac{\partial w({\bs{d}^{i,k}})}{\partial \theta_m} \frac{\partial w({\bs{d}^{i,k}})}{\partial \theta_n}  \left( X - \left( I_k w(\bs{d}^{i,k}) + B \right) \right)^2 \right)\right]\\
=& \frac{I_k^2}{\sigma _{noise}^4} \frac{\partial w({\bs{d}^{i,k}})}{\partial \theta_m} \frac{\partial w({\bs{d}^{i,k}})}{\partial \theta_n}
E\left[ \left( X - \left( I_k w(\bs{d}^{i,k}) + B \right) \right)^2 \right]\\
=& \frac{I_k^2}{\sigma _{noise}^2} \frac{\partial w({\bs{d}^{i,k}})}{\partial \theta_m} \frac{\partial w({\bs{d}^{i,k}})}{\partial \theta_n}
\end{aligned}
\end{equation}

The above derivation will hold for any distribution with zero-mean Gaussian noise.

\end{document}